\documentclass[12pt,a4paper]{article}

\usepackage{graphics}
\usepackage{epsfig}     

\usepackage[dvips]{color}
\usepackage{color}
\usepackage{amssymb}       
\usepackage{times}     
\usepackage{bm}

\begin{document}

\author{
Ralf Steuer$~^\mathrm{a,c}$\footnote{steuer@agnld.uni-potsdam.de}, Thilo Gross$^\mathrm{a}$, Joachim Selbig$^\mathrm{b,c}$, and Bernd Blasius$^\mathrm{a}$ \vspace{1cm} \\ 
{\small \em $^\mathrm{a}$University Potsdam, Institute for Physics, Nonlinear Dynamics Group,} \\
{\small \em                 Am Neuen Palais~10, 14469~Potsdam, Germany } \\
{\small \em $^\mathrm{b}$Max-Planck-Institute for Molecular Plant Physiology,} \\ 
{\small \em Am M\"uhlenberg 1, 14476 Potsdam-Golm, Germany }\\
{\small \em $^\mathrm{c}$University Potsdam, Institute for Biochemistry and Biology,} \\
{\small \em  Karl-Liebknecht-Strasse 24-25, Haus 20, 14476 Potsdam, Germany }
}

\title{Structural Kinetic Modeling of Metabolic Networks}
\date{\today}

\maketitle

\abstract{ \bf
 To develop and investigate detailed mathematical models of cellular metabolic 
 processes is one of the primary challenges in systems biology.
 However, despite considerable advance
 in the topological analysis of metabolic networks,
 explicit kinetic modeling based on differential equations is still 
 often severely hampered by inadequate knowledge of the 
 enzyme-kinetic rate laws and their associated parameter values.
 Here we propose a method that aims to give a detailed and quantitative account of
 the dynamical capabilities of metabolic systems, 
 without requiring 
 any explicit information about the particular functional form of the 
 rate equations.
 Our approach is based on constructing 
 a local linear model at each point in parameter space, 
 such that each element of the model
 is either directly experimentally accessible, 
 or amenable to a straightforward biochemical interpretation.
 This ensemble of local linear models, encompassing all possible explicit kinetic models,
 then allows for a systematic statistical exploration 
 of the comprehensive parameter space. 
 The method is applied to two paradigmatic examples: 
 The glycolytic pathway of yeast and a realistic-scale representation 
 of the photosynthetic Calvin cycle. 
}

\small
\section*{Introduction}
Cellular metabolism constitutes a complex dynamical system and gives 
rise to a wide variety of dynamical phenomena, including multistability and temporal oscillations.
To elucidate, understand and eventually predict the behavior of
metabolic systems 
represents one of the primary challenges in the postgenomic era~\cite{Palsson00,Kell_2004,FTKW04,WP04}.
To this end, substantial effort was dedicated in the recent years 
to develop and investigate detailed kinetic models of cellular metabolic processes~\cite{HS96,THT99}. \\  
Once a mathematical model is established, it can serve a multitude of purposes:
It can be regarded as a ``{\em virtual laboratory}'' that allows to build up a 
characteristic description of the system,
irrespective of experimental
restrictions, and gives insights into 
fundamental design principles of cellular functions, such as adaptability, 
robustness and optimality. 
Important questions often concern the existence and size of regions in the space of parameters
with qualitatively different behavior, 
such as multiple steady states (multistability) or autonomous oscillations~\cite{MWBB02,SSS04,AFS04}.
Likewise, mathematical models of cellular metabolism serve as a basis to  
investigate questions of major biotechnological importance, such as the 
effects of directed modifications of enzymatic or regulatory activities to
improve a desired property of the system~\cite{SAM04}. \\
However, while there has been a formidable progress in the structural (or topological) analysis
of metabolic systems~\cite{SFD00,FFN03}, 
and despite the long history of metabolic modeling,
dynamic models of cellular metabolism incorporating a 
realistic complexity are still scarce. \\
This is owed to the fact that the construction of such models 
encompasses a number of profound difficulties.
Most importantly, the construction of kinetic models relies on the precise knowledge of the
functional form of all involved enzymatic rate equations and their associated parameter values.
Furthermore, even if both is available from the literature, parameter values may (and usually do)
depend on many factors such as tissue type, or experimental and physiological conditions.
Likewise, most enzyme-kinetic rate laws have been determined {\em in vitro} and often there is only 
little guidance available whether a particular rate function is still appropriate {\em in vivo}. \\
In this work, we aim to overcome some of these difficulties and 
propose a bridge between structural modeling, which is based on the stoichiometry alone~\cite{SFD00,FFN03,Bai01}, 
and explicit kinetic models of cellular metabolism. 
In particular, we demonstrate that it is possible to acquire an exact, detailed 
and quantitative picture of the bifurcation structure of a given metabolic system, 
without explicitely referring to any particular set of differential equations.  \\
Our approach starts with the observation that in most
circumstances an explicit kinetic model is not necessary.
For example, to determine under which conditions 
a steady state looses its stability, only a local linear approximation of the 
system at this respective state is needed, i.e.  
we only need to know the eigenvalues of the associated Jacobian matrix. 
Note that by saying this, and unlike related approaches to qualitative modeling~\cite{GC05,Bai01},
we do not aim at an approximation of the system. 
The boundaries of an oscillatory region in parameter space that arise
out of a Hopf bifurcation are actually and exactly determined by 
the eigenvalues of the Jacobian. 
Likewise, other bifurcations, including bifurcations of higher codimension, can
be deduced from the spectrum of eigenvalues and give rise to specific dynamical behavior. \\
The basis of our approach thus consists of giving a parametric representation
of the Jacobian matrix of an arbitrary metabolic system at each possible point in parameter space,
such that each element is accessible even without explicit 
knowledge of the functional form of the rate equations.
Once this parametric representation of the Jacobian is obtained, 
it allows to give a detailed statistical account 
of the dynamical capabilities of a metabolic system, 
including the 
stability of steady states, 
the possibility of sustained oscillations, as well as the 
existence of quasiperiodic and chaotic regimes.
Moreover, the analysis is quantitative, i.e. it allows to deduce
specific biochemical conditions under which a certain dynamical behavior 
occurs and allows to assess the plausibility or robustness 
of experimentally observed behavior by
relating it to a quantifiable region in parameter space.

\section*{Structural Kinetic Modeling}
The temporal behavior of an arbitrary  metabolic reaction network, consisting
of $m$ metabolites and $r$ reactions,  can be
described by a set of differential equations~\cite{HS96},  
\begin{equation} \label{eq:Metabolism}
\frac{\mathrm{d}\mathbf{S}(t)}{\mathrm{d}t} = \mathbf{N} \, \bm{\nu}(\mathbf{S},\mathbf{k}) 
\end{equation}
where $\mathbf{S}$ denotes the $m$-dimensional vector of biochemical reactants and
$\mathbf{N}$ the $m \times r$ stoichiometric matrix. 
The $r$-dimensional vector of reaction rates $\bm{\nu}(\mathbf{S},\mathbf{k})$ consists 
of nonlinear (and often unknown) functions,
which depend on the substrate concentrations $\mathbf{S}$, 
as well as on a set of (often unknown) parameters $\mathbf{k}$. \\
In the following, we will not assume explicit knowledge of the functional
form of the rate equations, but instead aim at a parametric 
representation of the Jacobian of the system.
As the only mathematical assumption about the system, we require
the existence of a positive state $\mathbf{S}^0$ that fulfills the steady state 
condition $\mathbf{N} \bm{\nu}(\mathbf{S}^0,\mathbf{k})=\mathbf{0}$. 
Importantly, the state $\mathbf{S}^0$ is neither required to be unique, nor stable. \\
Using the definitions
\begin{equation} \label{eq:rates}
x_i(t) := \frac{S_i(t)}{S_i^0}, 
\quad 
\Lambda_{ij} := N_{ij}   \frac{\nu_j(\mathbf{S^0})}{S_i^0}
\,\,\, \mbox{and} \,\,\, \mu_j(\mathbf{x}) :=  \frac{\nu_j(\mathbf{x})}{\nu_j(\mathbf{S^0})} 
\end{equation}
and following the normalization procedure proposed in~\cite{Gross04,GF05},
the system can be straightforwardly rewritten in terms of new variables $\mathbf{x}(t)$
\begin{equation}
\frac{\mathrm{d}\mathbf{x}}{\mathrm{d}t} = \bm{\Lambda} \, \bm{\mu}(\mathbf{x}) 
\qquad \mbox{.}
\end{equation}
The corresponding Jacobian of the normalized system at the steady state $\mathbf{x}^0=\mathbf{1}$ is
\begin{equation} \label{eq:Jacobian}
\mathbf{J}_\mathbf{x} = \bm{\Lambda} \, 
\left. \frac{\partial \bm{\mu}(\mathbf{x})}{\partial \mathbf{x}} \right|_{\mathbf{x}^0=\mathbf{1}}
=: \bm{\Lambda} \, \bm{\theta}_\mathbf{x}^\mathbf{\bm \mu}
\end{equation}
As the new variables $\mathbf{x}$ are related to $\mathbf{S}$ by a simple
multiplicative constant, $\mathbf{J}_\mathbf{x}$ can be straightforwardly transformed 
back into the original Jacobian. \\
Any further discussion now rests crucially on the interpretation of the
terms in Eq.~(\ref{eq:Jacobian}). Once these coefficients are known, 
the Jacobian of the system can be evaluated. 
We begin with an analysis of the matrix $\bm{\Lambda}$.
Its elements $\Lambda_{ij}$ have the units of an inverse time and
consist of the elements of the stoichiometric matrix $\mathbf{N}$,
the vector of steady state concentrations $\mathbf{S}^0$, a
nd the steady state fluxes $\bm{\nu}(\mathbf{S}^0)$.
Provided a metabolic system is designated for mathematical modeling, 
we can safely assume that there exists some knowledge about
the relevant concentrations,  
i.e. for each metabolite, we can specify an interval 
$S_i^- \;  \leq  \;  S_i^0 \;  \leq \;  S_i^+$
which defines a physiologically feasible range of the respective concentration.
Furthermore, the steady state fluxes $\bm{\nu}(\mathbf{S}^0)$ are subject to the 
mass-balance constraint $\mathbf{N}\bm{\nu}(\mathbf{S}^0) = \mathbf{0}$, leaving only
$r-\mathrm{rank}(\mathbf{N})$ independent reaction rates~\cite{HS96}. 
Again, an interval $\nu_i^- \leq \nu_i^0 \leq \nu_i^+$ can be 
specified for all independent reaction rates, defining a physiologically admissible flux-space. \\
In the following, we denote $\mathbf{S}^0$ and $\bm{\nu}(\mathbf{S}^0)$, usually 
corresponding to an experimentally observed state of the system, 
as the {\em operating point} at which the Jacobian is to be evaluated. 
This information, together with the stoichiometric matrix $\mathbf{N}$, fully specifies 
the matrix $\bm{\Lambda}$. \\
The interpretation of the matrix $\bm{\theta}_\mathbf{x}^\mathbf{\bm \mu}$ 
in Eq.~(\ref{eq:Jacobian}) is slightly more subtle
since it involves the derivatives of the unknown functions $\bm{\mu}(\mathbf{x})$ 
with respect to the new normalized variables $\mathbf{x}$ at the point $\mathbf{x}^0 = \mathbf{1}$. 
Nevertheless, an interpretation of these parameters is possible and
does not rely on the explicit knowledge of the detailed functional form of the rate equations:
Each element $\theta_{x_i}^{\mu_j}$ of the matrix $\bm{\theta}_\mathbf{x}^\mathbf{\bm \mu}$ measures 
the normalized degree of saturation 
of the reaction $\nu_j$ with respect to a substrate $S_i$ at the operating point $\mathbf{S^0}$. \\
In particular, the dependence of almost all biochemical rate laws $\nu_j(\mathbf{S})$ 
on a biochemical reactant $S_i$ 
can be written in the form $\nu_j(\mathbf{S},\mathbf{k})=k_\mathrm{v} S_i^n/f_m(\mathbf{S},\mathbf{k})$, where
$f_m(\mathbf{S},\mathbf{k})$ denotes a polynomial of order $m$ in $S_i$ with positive coefficients $\mathbf{k}$. 
All other reactants have been absorbed into $\mathbf{k}$~\cite{HS96}. 
After applying the transformation of Eq.~(\ref{eq:rates}), 
we obtain 
\begin{equation} \label{eq:GeneralRate}
\theta^{\mu_j}_{x_i} = 
\left. \frac{\partial\mu_j(\mathbf{x})}{\partial x_i} \right|_{\mathbf{x}^0=\mathbf{1}}  = n - \alpha \, m
\end{equation}
with $\alpha \in [0,1]$ denoting a free variable in the unit interval.
The limiting cases are {\em always} $\lim_{S^0_i\rightarrow 0} \alpha= 1$ and $\lim_{S^0_i\rightarrow \infty} \alpha= 0$.
To evaluate the matrix $\bm{\theta}_\mathbf{x}^\mathbf{\bm \mu}$  we thus 
restrict each saturation parameter  
to a well-defined interval, specified in the following way:
As for most biochemical rate laws $n=m=1$, the partial derivative usually takes a value 
between zero and unity, determining the degree of saturation of the respective reaction. 
In the case of cooperative behavior with a Hill coefficient $n = m \geq 1$, 
the normalized partial derivative lies in the interval $[0,n]$  and, analogously, 
in the interval $[0,-m]$ for inhibitory interaction with $n=0$ and $m \geq 1$. 
For examples and proof of Eq.~(\ref{eq:GeneralRate}) see {\em Materials and Methods}. \\
The matrices $\bm{\theta}_\mathbf{x}^\mathbf{\bm \mu}$ and $\bm{\Lambda}$, defined above,
now fully specify the Jacobian of the system.  
In the following, both quantities are treated as free parameters, 
defining the physiologically admissible {\em parameter space} of the system.  
Importantly, our representation of the Jacobian fulfills three essential conditions:
{\em i)} The reconstructed Jacobian represents the exact Jacobian at this
point in parameter space. There is no approximation involved.
{\em ii)} Each term in the Jacobian is either directly experimentally accessible, 
such as flux or concentration values, or has a well-defined biochemical 
interpretation, such as a normalized degree of saturation of a given reaction. 
{\em iii)} The Jacobian does not depend on any particular choice 
of specific rate functions. Rather, it encompasses {\em all} possible 
kinetic models of the system that are consistent with the considerations above.
In particular, any specific kinetic model, involving a specific choice of biochemical rate functions, 
can be mapped onto a particular point or region of the generalized parameter space. 
In this sense, the reconstructed Jacobian is exhaustive.

\subsection*{An illustrative Example}
\begin{figure}
\centering{\includegraphics[width=0.3\textwidth]{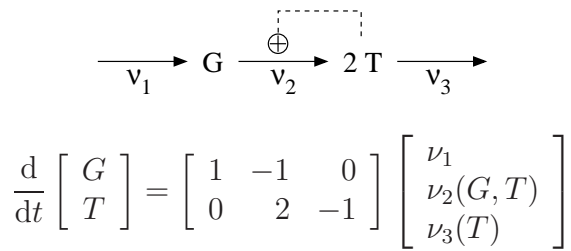}}
\begin{displaymath}
\frac{\mathrm{d}}{\mathrm{d}t} 
\left[ \begin{array}{c} G \\ T  \end{array}    \right] =
\left[ \begin{array}{rrr} 1 & -1 & 0 \\ 0 & 2 & -1  \end{array}    \right]
 \left[ \begin{array}{l} \nu_1 \\ \nu_2(G , T) \\ \nu_3(T)  \end{array}    \right] 
\end{displaymath}
\caption{\label{fig:BierModel} \small
A simple hypothetical pathway, reminiscent of a minimal model of yeast glycolysis~\cite{BBW00}: 
One unit of glucose (G) is converted into two units of ATP (T),
with ATP exerting a positive feedback on its own production.
The lower panel depicts the corresponding set of differential equations with the as yet
unspecified rate functions $\nu_1=const, \nu_2(G,T)$ and $\nu_3(T).$
}
\end{figure}

Prior to an application to more detailed biochemical models, we
exemplify our approach using a simple hypothetical pathway.
Suppose the reaction scheme depicted in Fig.~\ref{fig:BierModel},
consisting of  $m=2$ metabolites and $r=3$ reactions, 
is designated for mathematical modeling. 
The starting point of our analysis is then a particular
operating point, characterized by the metabolite
concentrations $\mathbf{S}^\mathbf{0}=(G^0,T^0)$ and 
flux values $\bm{\nu}^\mathbf{0}=(\nu_1^0,\nu_2^0,\nu_3^0)$. 
Within a reasonably realistic scenario, we can assume that  
the average concentrations of both metabolites have been determined experimentally.
Furthermore, an analysis of the stoichiometric matrix $\mathbf{N}$ 
reveals that there is only one independent steady state reaction rate $c$, 
with $\nu_1^0 = \nu_2^0 =c$ and $\nu_3^0=2c$.
Thus we only require knowledge
of the average overall flux through the pathway, specifying the value $c$. \\
This information already enables the construction of the matrix $\bm{\Lambda}$, 
which defines the (usually experimentally observed) operating point 
at which the system is to be evaluated.
\begin{equation}
\bm{\Lambda} = \left[ 
\begin{array}{ccc} 
c/G^0  & -c/G^0  & 0 \\
   0   & 2c/T^0  & -2c/T^0 \\
 \end{array} \right] 
\end{equation}
The only remaining parameters are now the elements of 
the matrix $\bm{\theta}_\mathbf{x}^\mathbf{\bm \mu}$.
Starting with the dependence of each reaction upon its substrate and
assuming conventional biochemical rate laws, we obtain
$\theta^{\mu_2}_\mathrm{G}\in [0,1]$, specifying the  degree of 
saturation of $\nu_2$ with respect to its substrate G.
Furthermore, $\theta^{\mu_3}_\mathrm{T}\in [0,1]$ specifies the degree of 
saturation of $\nu_3$ with respect to T.
Additionally, the known regulatory feedback of the metabolite T upon the reaction~$\nu_2$ is 
incorporated by $\theta^{\mu_2}_\mathrm{T}\in [0,n]$, 
where $n \geq 1$ denotes a positive integer (Hill coefficient). 
The matrix $\bm{\theta}_\mathbf{x}^\mathbf{\bm \mu}$ thus contains three  
nonzero values, each restricted to a well-defined interval
\begin{equation}
\bm{\theta}_\mathbf{x}^\mathbf{\bm \mu} = 
\left[
\begin{array}{cc}
0 & 0 \\
\theta^{\mu_2}_\mathrm{G} & \theta^{\mu_2}_\mathrm{T} \\
0         & \theta^{\mu_3}_\mathrm{T}  
 \end{array} \right] ~~.
\end{equation}
We emphasize that the three elements of $\bm{\theta}_\mathbf{x}^\mathbf{\bm \mu}$ 
represent {\em bona fide} parameters of the system,
specifying the Jacobian matrix   
no less unique and quantitative than 
a corresponding set of Michaelis constants, 
albeit without referring to the explicit functional form of any rate equation.  
Given the elements of $\bm{\theta}_\mathbf{x}^\mathbf{\bm \mu}$ 
as adjustable parameters, we 
have thus obtained a parameteric representation of the Jacobian matrix
which encompasses all possible kinetic models consistent with the experimentally
observed operating point.
In the remainder of this paper, 
we utilize our approach to evaluate the dynamical capabilities of two more
complex examples of metabolic system.

\section*{The Glycolytic Pathway}
\begin{figure}
\centering{\includegraphics[width=0.49\textwidth]{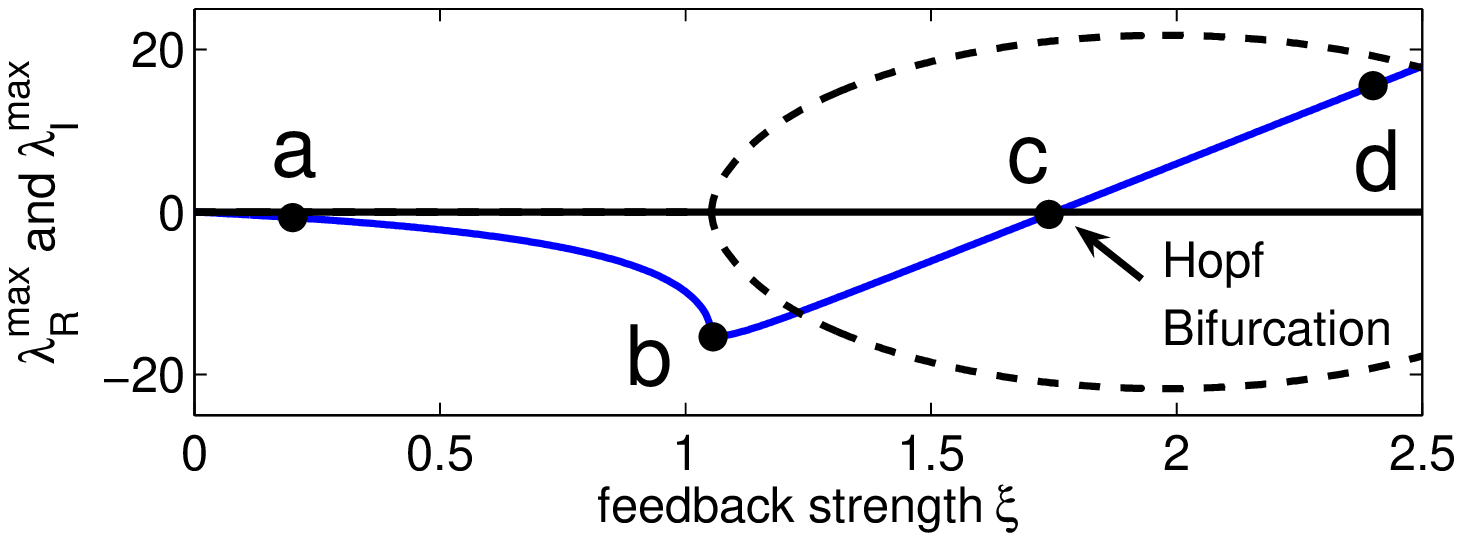}}
\centering{
\resizebox{0.5\textwidth}{!}{
           \includegraphics[width=0.12\textwidth]{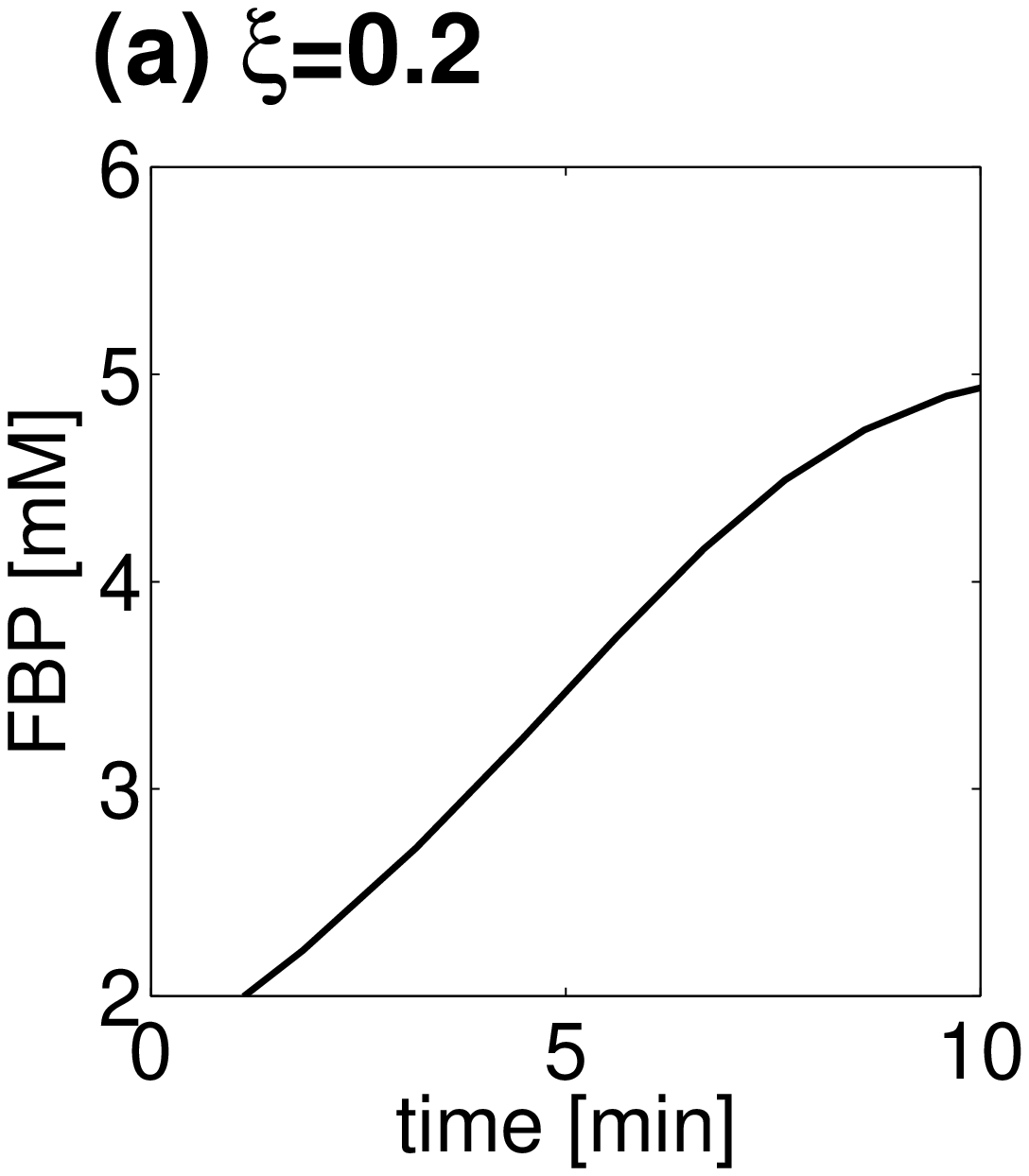}
           \includegraphics[width=0.12\textwidth]{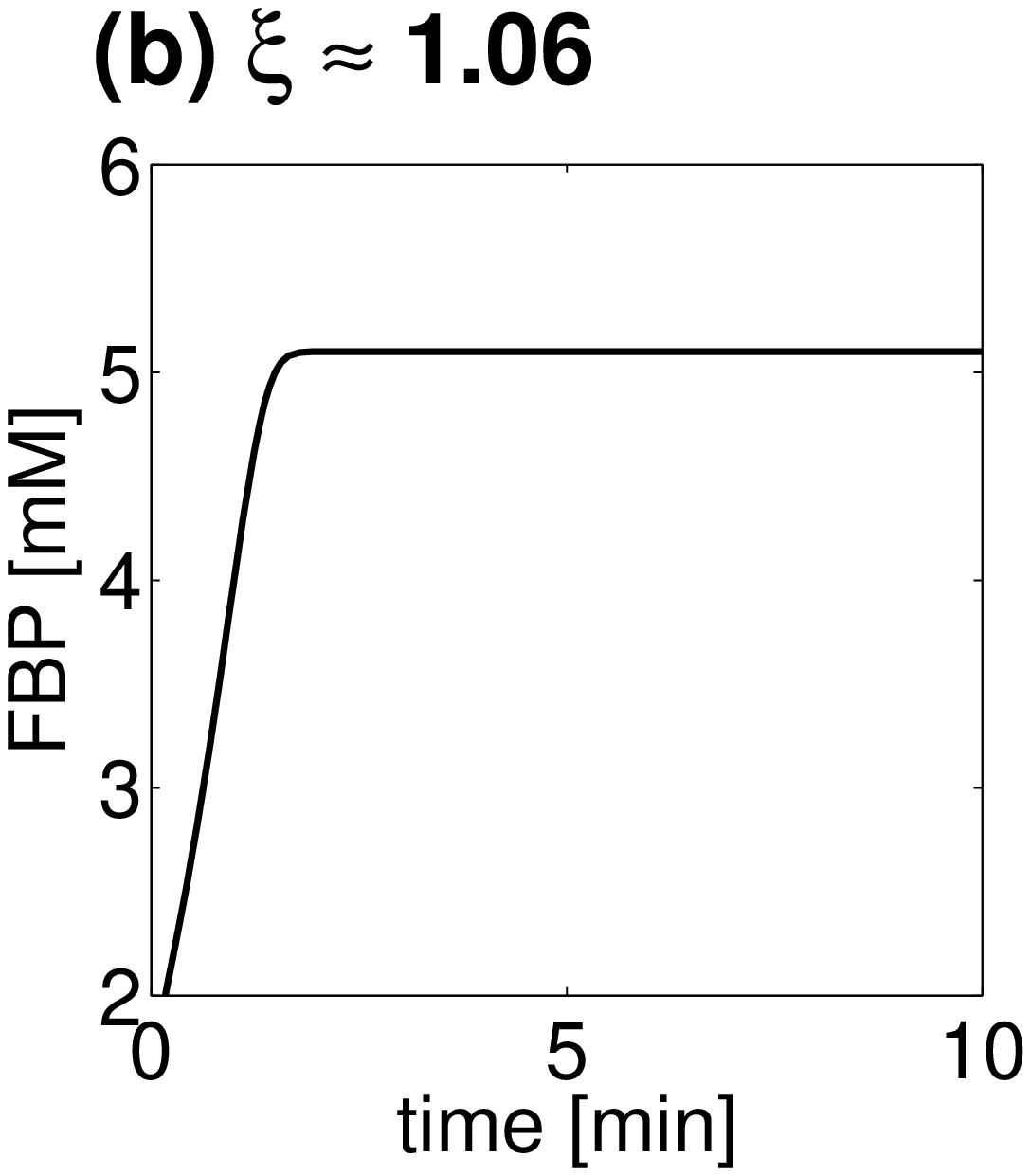}
           \includegraphics[width=0.12\textwidth]{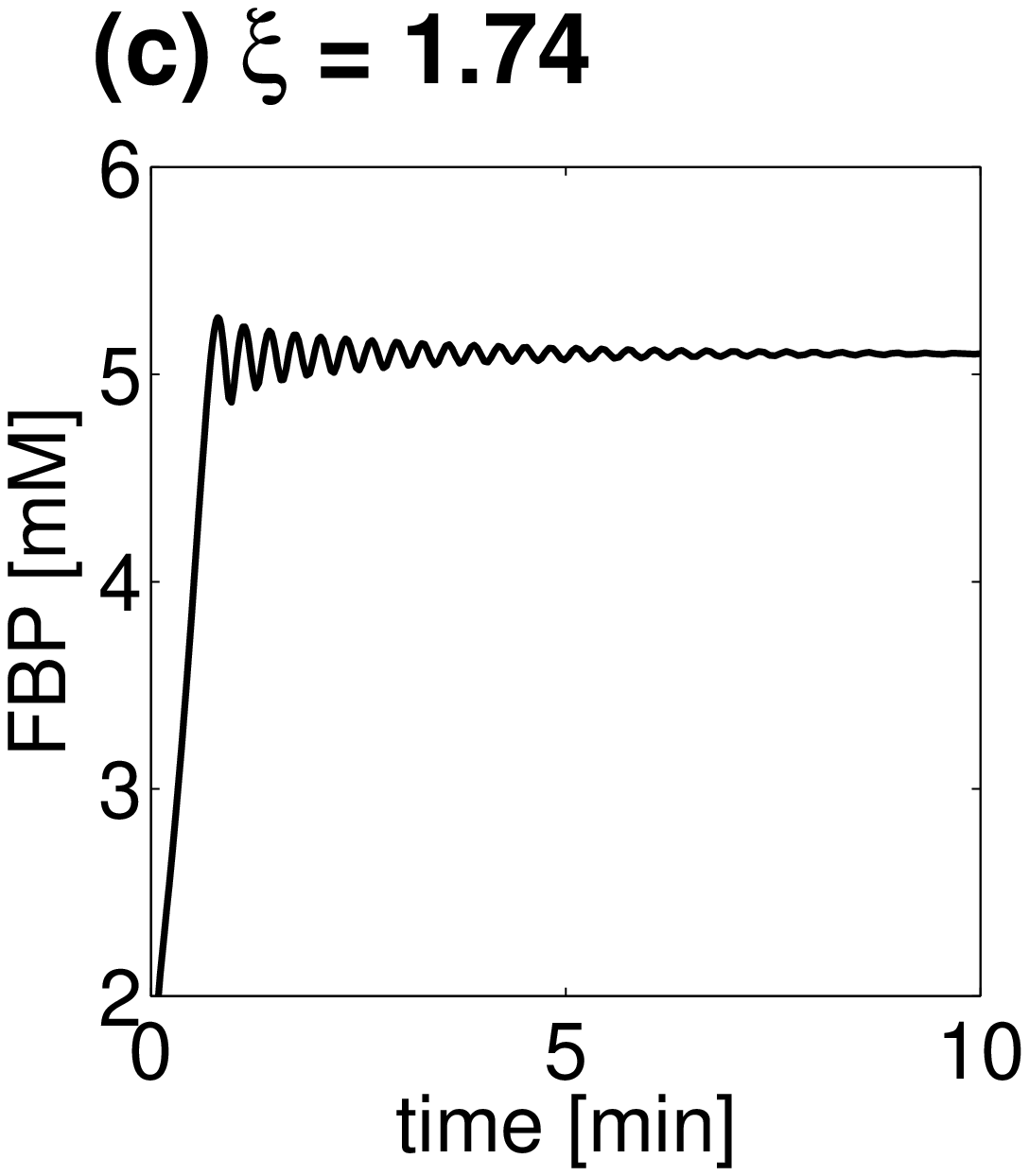}
           \includegraphics[width=0.12\textwidth]{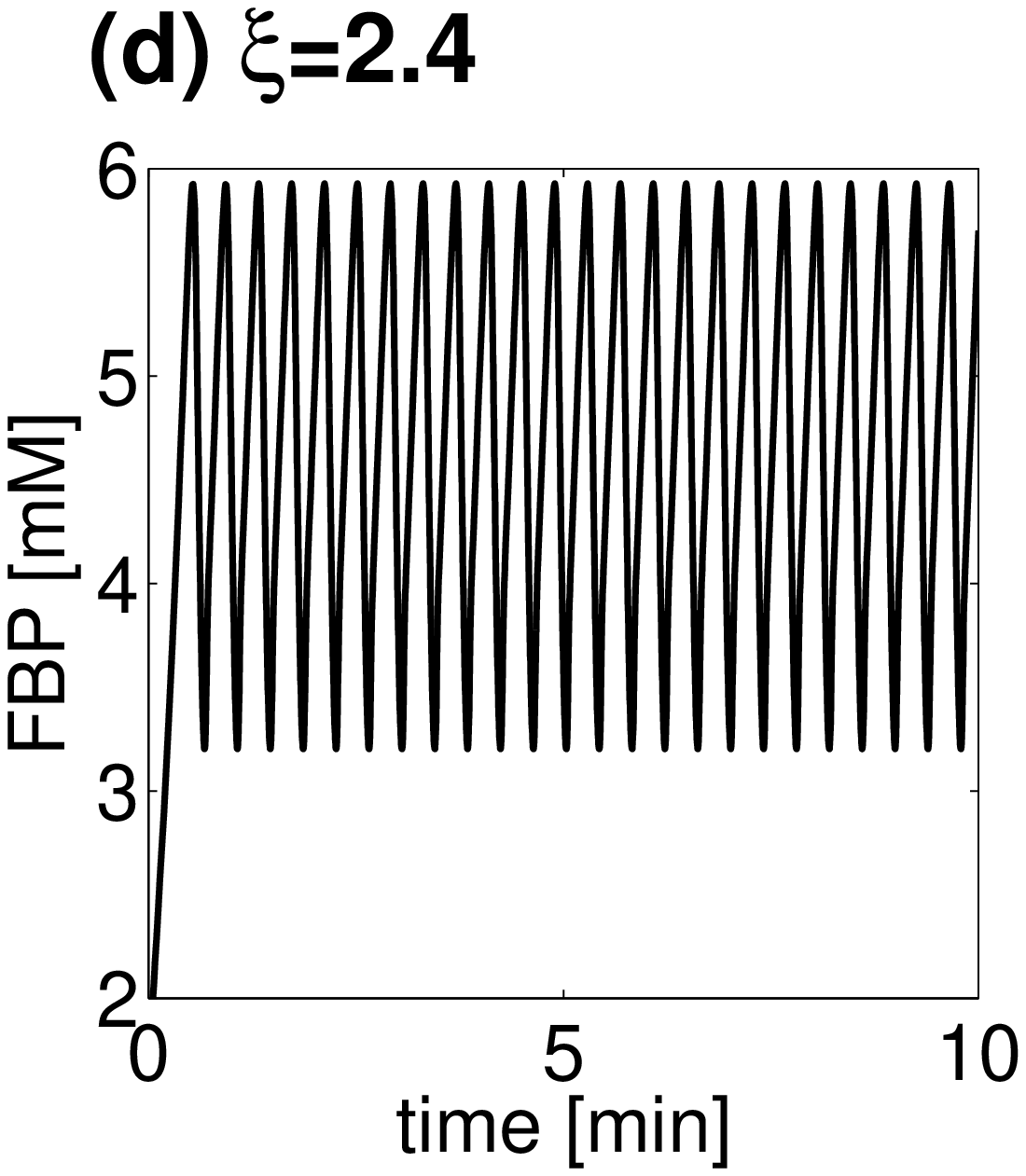}
}}
 \caption{\label{fig:LargestEV} \small
{\em Upper plot:} The eigenvalue with the largest real part 
as a function of the inhibitory feedback strength $\xi$ of ATP on the combined PFK-HK reaction~$\nu_1$.
All other saturation parameter are $\theta^\mu_x=1$. 
Shown is the real part $\lambda_R^\mathrm{max}$ (solid line) together 
with the imaginary part $\lambda^\mathrm{max}_I$ (dashed line). 
At the Hopf bifurcation a complex conjugate pair of eigenvalues 
$\lambda^\mathrm{max}=\lambda_R^\mathrm{max} \pm i \lambda^\mathrm{max}_I$
crosses the imaginary axis. 
{\em Lower plots:} 
A varying feedback strength $\xi$ allows for $4$ different dynamical regimes.
Shown are time courses of FBP using an an explicit kinetic model 
at the points (a,b,c,d) indicated above.  
{\em a)} Slow relaxation to the stable steady state.  
{\em b)} Optimal response to perturbations, as determined by the minimal 
largest eigenvalue $\lambda_R^\mathrm{max}$.
{\em c)} Oscillatory return to the stable steady state.
{\em d)} Sustained oscillations.
All different regimes can be deduced solely from the Jacobian 
and are only exemplified using the explicit kinetic model.
For rate equations and kinetic parameters see Supplementary Information.
}
\end{figure}

Among the most classical and probably best studied example of a 
biochemical oscillator is the breakdown of sugar via glycolysis in yeast.
Damped and sustained glycolytic oscillations have been observed for several decades 
and have triggered the development of a large variety of
kinetic models, ranging from simple minimal models~\cite{BBW00} 
to more elaborate representations of the pathway~\cite{WPSS00,HDS01}. \\ 
In the following, we will address some of the characteristic questions that
led to the development of those earlier models, and show that these can
be readily answered using the concept of  structural kinetic modeling. 
\begin{figure}
\centering{ \includegraphics[width=0.49\textwidth]{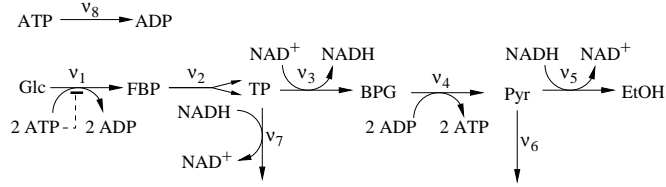}}
 \caption{\label{fig:Glycolysis} \small
A medium-complexity representation of the yeast glycolytic pathway,
adapted from earlier kinetic models~\cite{WPSS00}.
The system consists of $8$ metabolites and $8$ reactions.
The crucial regulatory step is the phosphofructokinase (PFK), here 
combined with the hexokinase (HK) reaction into the reaction rate $\nu_1$. 
Though PFK is known to have several effectors, we only consider the inhibition by
its substrate ATP, again following earlier kinetic models~\cite{WPSS00}. 
For metabolite abbreviations and the observed metabolite concentration  
see the Supplementary Information accompanying this article.
}
\end{figure}
Given a schematic representation of the pathway, as depicted in Fig.~\ref{fig:Glycolysis},
the first and foremost problem is to establish whether the proposed
reaction mechanism indeed facilitates  
sustained oscillations at the experimentally observed operating point. 
And, if yes, what are the specific kinetic conditions and requirements 
under which sustained oscillations can be expected.   \\
We start out by constructing the matrix $\bm{\Lambda}$
using the experimentally observed state $\mathbf{S}^0$ and $\bm{\nu}^0$,
identified here with the average concentration and flux values reported in~\cite{WPSS00,HDS01}. 
Furthermore, the matrix of saturation coefficients $\bm{\theta}_\mathbf{x}^\mathbf{\bm \mu}$
has to be specified.
For simplicity, we assume that all reactions are 
irreversible and depend on their respective substrates only, resulting
in $13$ free parameters.
Based on our discussion of conventional biochemical rate laws above, 
the saturation coefficients are restricted to the unit interval $\theta_{S}^{\mu} \in [0,1]$.  \\
For the dependence of the PFK-HK reaction on ATP, we follow a previously proposed
kinetic model~\cite{WPSS00} and assume a linear activation
due to its effect as a substrate and a saturable inhibition involving a 
positive Hill coefficient $n \geq 1$. 
The corresponding parameter is 
thus $\theta^{\mu_1}_\mathrm{ATP}=1-\xi$, with $\xi \in [0,n]$.
No further assumptions about the detailed functional form of 
any of the rate equations are necessary. 
For an explicit representation of both matrices $\bm{\Lambda}$ and 
$\bm{\theta}_\mathbf{x}^\mathbf{\bm \mu}$ see the Supplementary Information. \\
To investigate the possibility of sustained oscillation, we 
begin with the most simplest scenario and set $\theta_{S}^{\mu}=1$ for all reactions, 
corresponding to bilinear mass-action kinetics. 
Note, however, that the inhibition term is still assumed to be an unspecified nonlinear saturable function. 
Fig.~\ref{fig:LargestEV} shows the largest eigenvalue of the resulting Jacobian at the 
experimentally observed operating point as a function of the feedback strength $\xi$.
Indeed for sufficient inhibition, the spectrum of eigenvalues passes 
through a Hopf bifurcation and the system facilitates sustained oscillations.
Importantly, for a Hopf bifurcation to occur at the observed operating point, a Hill
coefficient $n \geq 2$ is needed, irrespective of the detailed functional form of the rate equation. \\
We have to highlight one fundamental aspect of our analysis:
Given our parametric representation of the Jacobian, the impact of the inhibition is
decoupled from the steady state concentrations and 
flux values the system adopts (the latter being solely determined by the matrix $\bm{\Lambda}$).
Thus, with Fig.~\ref{fig:LargestEV}, we specifically ask whether the assumed
inhibition is indeed a necessary condition for the 
observation of oscillations {\em at the experimentally observed operating point}.
In contrast to this, using a conventional kinetic model and reducing the influence of the regulation, i.e. 
by increasing the corresponding Michaelis constant, would concomitantly result in 
altered steady state concentrations -- thus not straightforwardly contributing to this question. \\  
Furthermore, as glycolytic oscillations have no obvious physiological role and are only
observed under rather specific experimental conditions, 
some questions concerning their possible functional significance have been raised.
One assertion is that the observed oscillations might 
only be an unavoidable side effect of the regulatory interactions, optimized for other purposes~\cite{HS96}.
Indeed, as shown in Fig.~\ref{fig:LargestEV}, a varying feedback strength $\xi$ allows for
different dynamical regimes. In particular, an intermediate value
speeds up the response time with respect to perturbations, as 
also frequently observed in explicit models of cellular regulation~\cite{REA02}.

\subsection*{Statistical Analysis of the Parameter Space}
\begin{figure}
\centering{\includegraphics[width=0.234\textwidth]{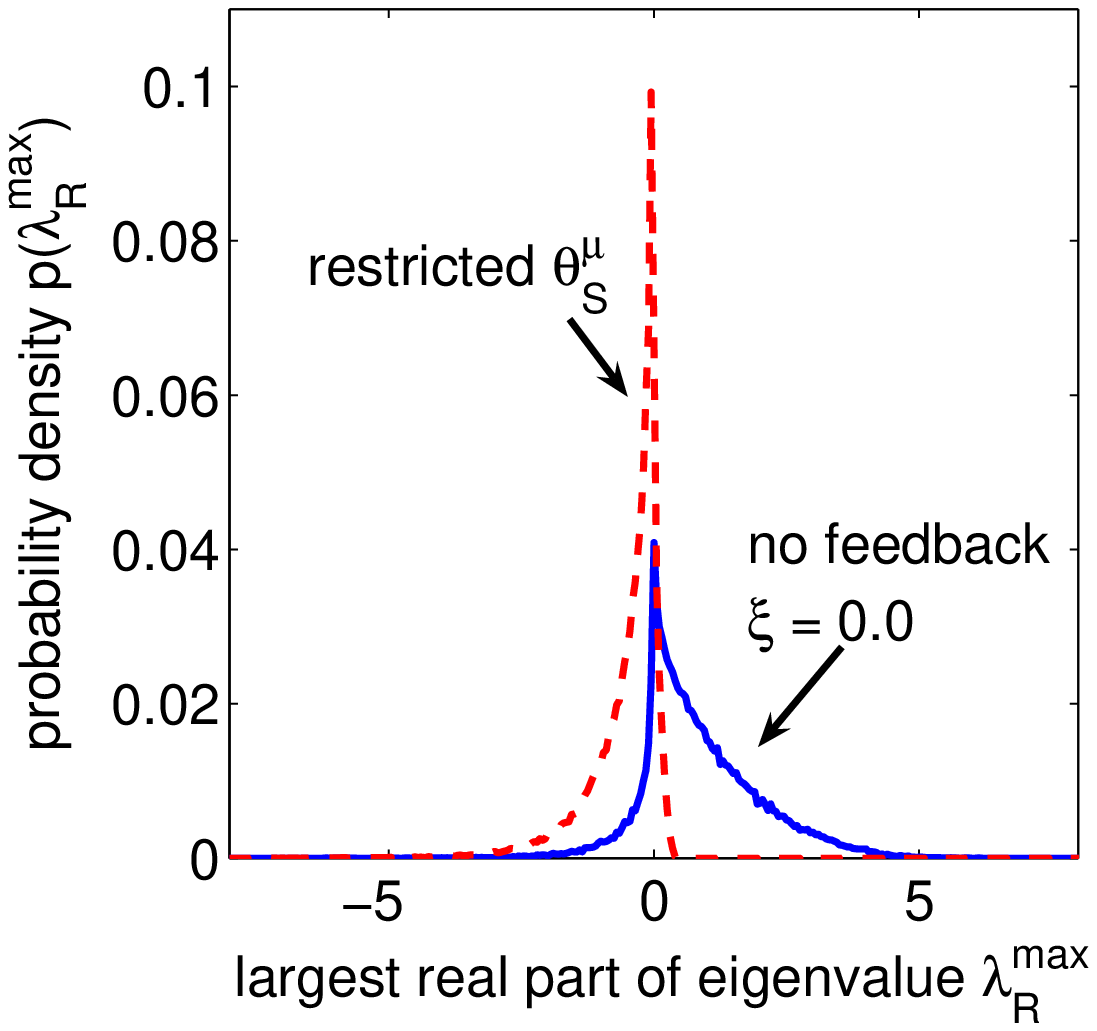}
           \includegraphics[width=0.230\textwidth]{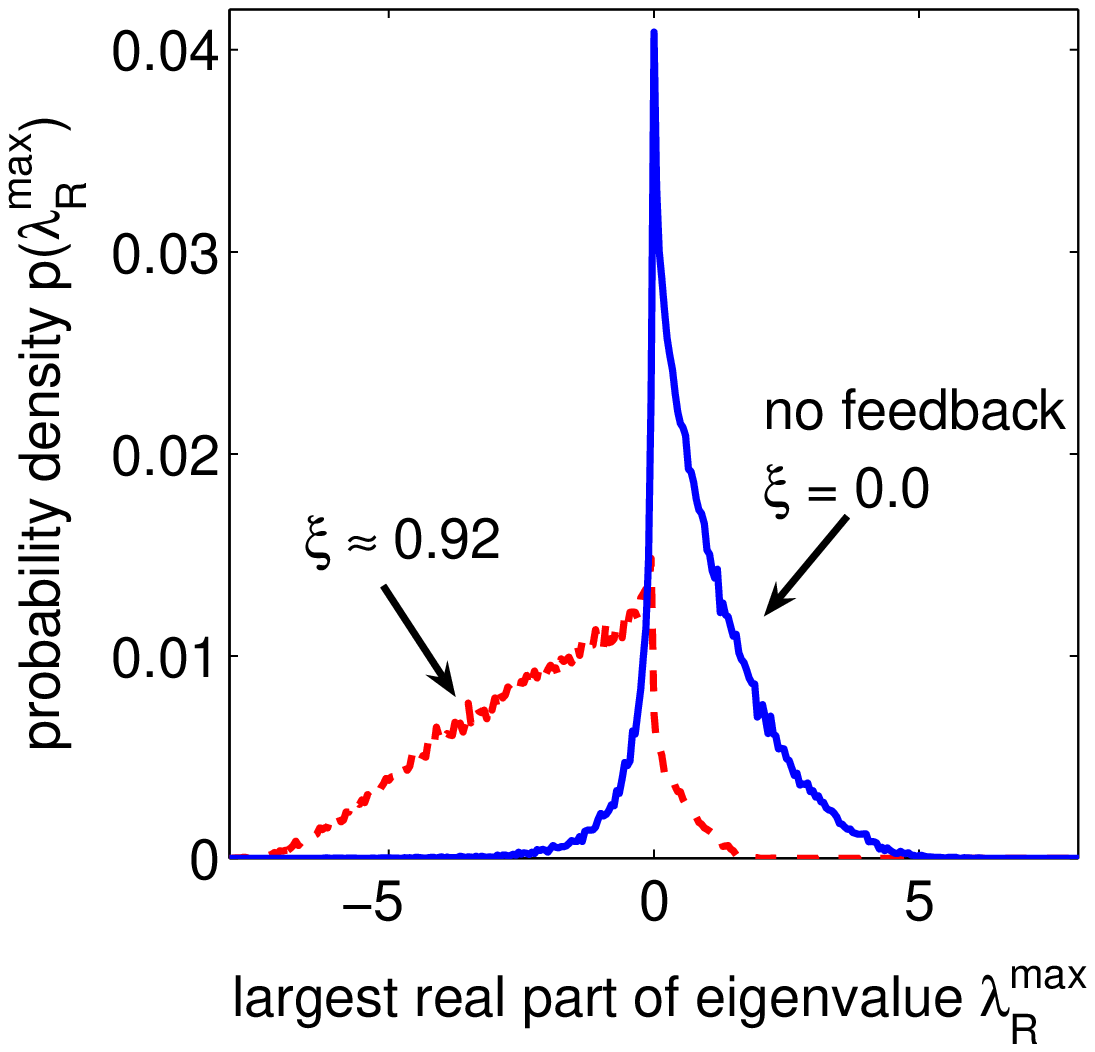}
}
 \caption{\label{fig:EV_histogram} \small
The distribution of the largest real part $\lambda^\mathrm{max}_R$ within the spectrum of eigenvalues 
for $10^5$ realizations of the Jacobian matrix at the operating point. 
For each realization, the $12$ saturation parameters $\theta^\mu_S$ were sampled randomly
from a uniform distribution in the unit interval. 
A value $\lambda^\mathrm{max}_R>0$ implies instability.
In the absence of the regulatory feedback ($\xi=0$, blue solid line), 
the observed operating point of the system
is likely to be unstable. 
{\em Left:} Influence of specific reaction rates. 
The saturation parameters showing the largest impact on $\lambda^\mathrm{max}_R$ 
are restricted to weak saturation: 
$\theta^{\mu_8}_\mathrm{ATP}=0.9$, $\theta^{\mu_6}_\mathrm{Pyr}=0.9$, and $\theta^{\mu_7}_\mathrm{TP}=0.9$.
The probability of finding $\lambda^\mathrm{max}_R<0$ is markedly increased (red dashed line). 
{\em Right:} For a nonzero feedback strength $\xi \approx 0.92$ the distribution of $\lambda^\mathrm{max}_R$
is shifted towards negative values, i.e. most realizations of the Jacobian give rise to a stable
steady state (red dashed line).
}
\end{figure}

\begin{figure}
\centering{\includegraphics[width=0.23\textwidth]{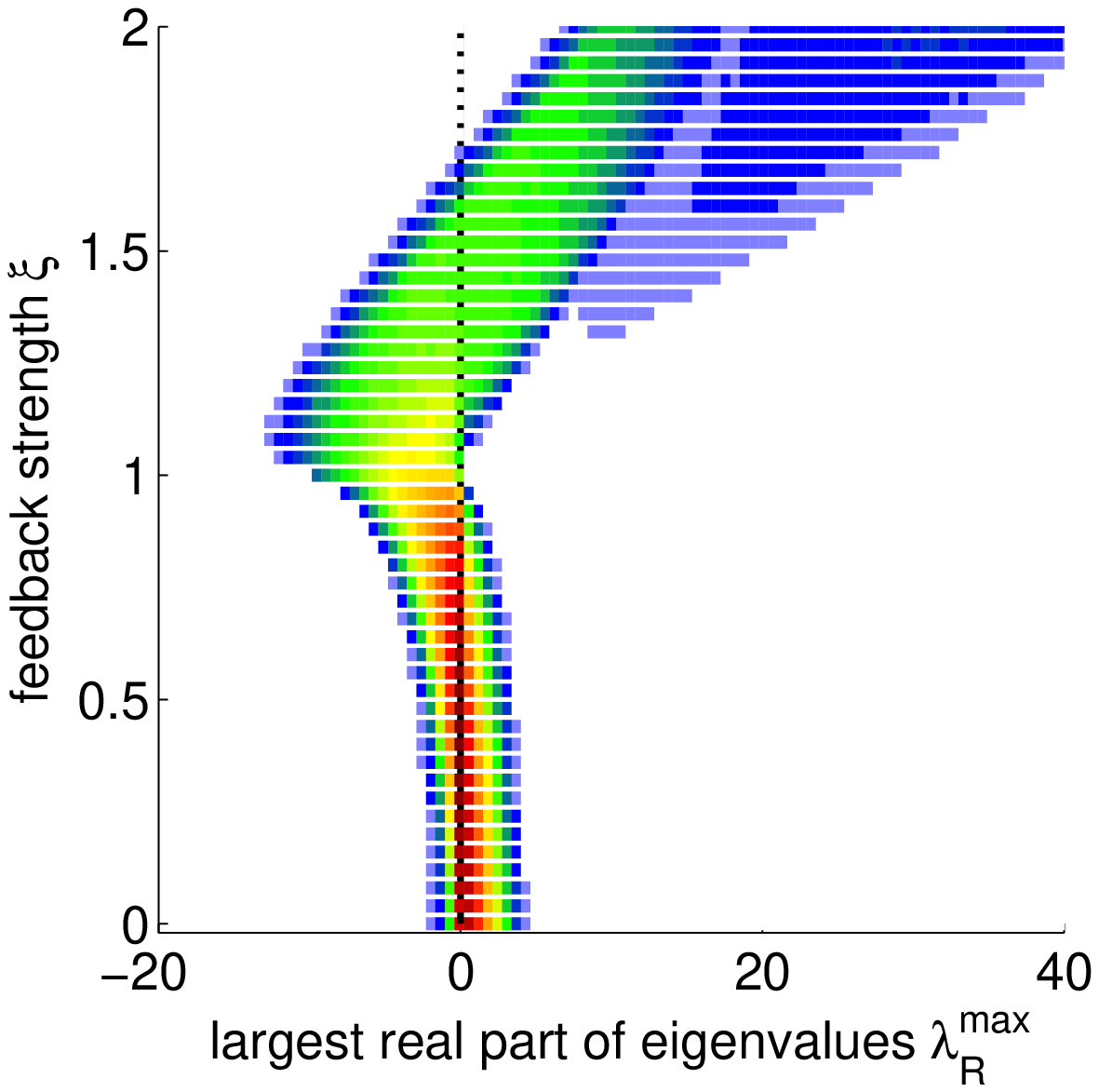}
           \includegraphics[width=0.23\textwidth]{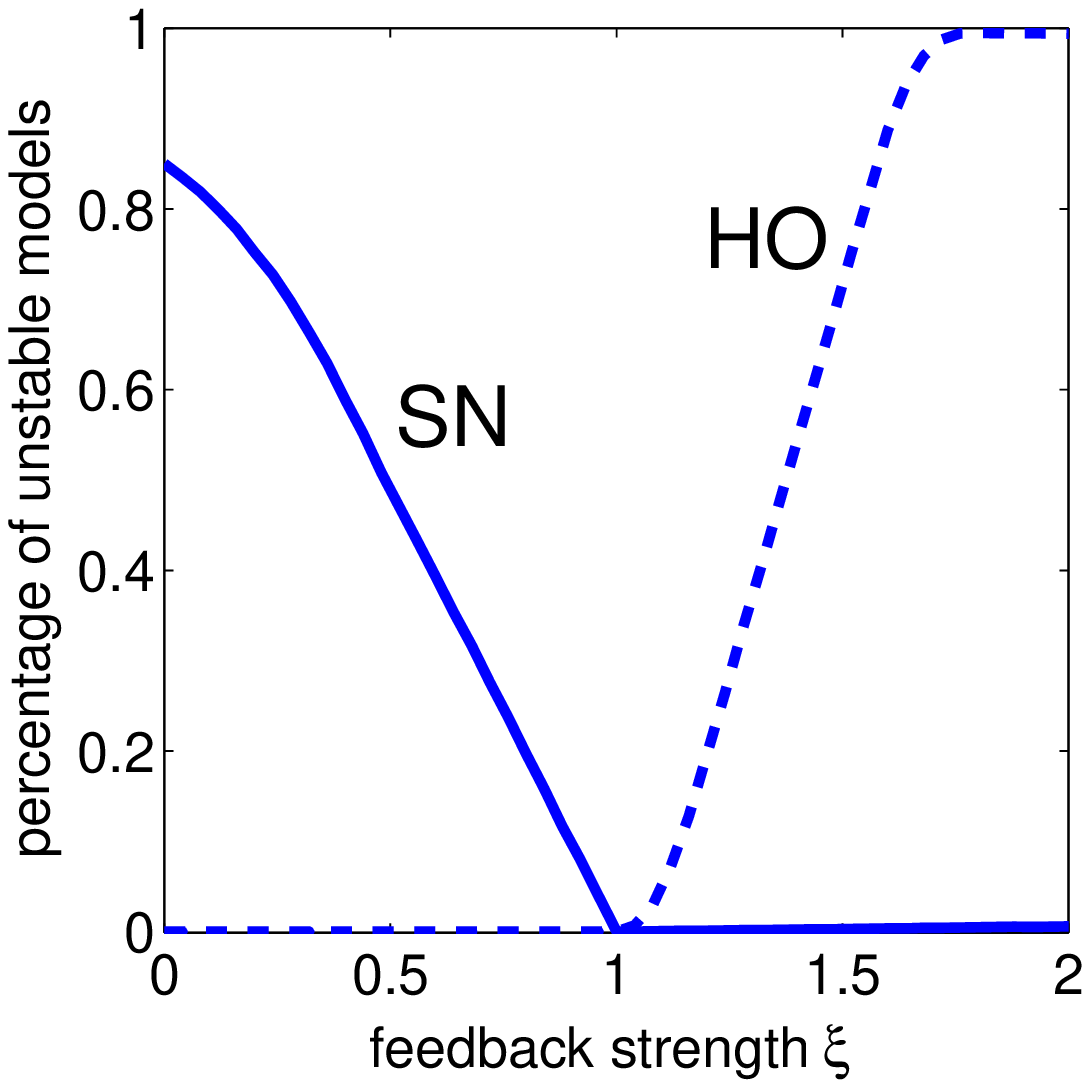}
}
 \caption{\label{fig:EV_Stability} \small
The distribution of the largest real part $\lambda^\mathrm{max}_R$ 
of the eigenvalues as a function of the feedback strength $\xi$. 
All other saturation parameters are sampled randomly from a uniform distribution.
{\em Left:} Color-coded visualization of the resulting distribution of~$\lambda^\mathrm{max}_R$ (red: large, blue: small). 
{\em Right:} The relative fraction of models with $\lambda^\mathrm{max}_R>0$, implying the 
instability of the observed operating point. 
The solid line (SN) denotes the case of a single positive real part $\lambda^\mathrm{max}_R$ within the spectrum of eigenvalues,
indicating a saddle-node bifurcation.
The dashed line (HO) denotes the case of a pair of complex conjugate eigenvalues with positive real parts.
Note that, strictly speaking, this does not necessarily imply sustained oscillations.
However, it indicates the existence of a nearby Hopf bifurcation, thus constituting prima facie evidence
for oscillatory behavior.
}
\end{figure}

Going beyond the case of bilinear kinetics, 
we now evaluate the properties of Jacobian at the most general level.
All saturation coefficients $\theta^\mu_S \in (0,1]$ are allowed 
to take arbitrary values in the unit interval, encompassing all possible explicit kinetic models 
of the pathway shown in Fig.~\ref{fig:Glycolysis}. 
The steady state concentrations and flux values are again restricted 
to the experimentally observed operating point. 
To assess the robustness of the system at this operating point,
the saturation coefficients $\theta^\mu_S \in (0,1]$ are 
repeatedly sampled from a uniform distribution. 
For each random realization the Jacobian is evaluated and 
the largest real part $\lambda^\mathrm{max}_R$ of its eigenvalues is recorded. 
Fig.~\ref{fig:EV_histogram} shows the histogram of the largest real part within
the spectrum of eigenvalues, with $\lambda^\mathrm{max}_R>0$ implying instability of the operating point.
In the absence of the inhibitory feedback $\xi=0$ the 
operating point is likely to be unstable, 
i.e. most realizations result in a spectrum of eigenvalues with at least one positive real part. \\
Two ways to circumvent this inherent instability are conceivable: 
First we can ask about the dependence on particular ~ reactions, that is,
whether the saturation (or non-saturation) of a specific reaction contributes to an 
increased stability of the system.
To this end, the correlation coefficient between $\lambda^\mathrm{max}_R$, reflecting the stability
of the system, and the saturation parameters $\theta^\mu_S$ was estimated. 
Indeed, several parameters $\theta^\mu_S$ show a strong correlation with $\lambda^\mathrm{max}_R$, 
indicating that their value essentially determines the stability of the system (for data see Supplementary Information).  
Fig.~\ref{fig:EV_histogram}a depicts the distribution of $\lambda^\mathrm{max}_R$ under the assumption that 
these reactions are restricted to weak saturation. 
In this case, the resulting distribution is shifted towards negative values, 
corresponding to an increased probability of the system to operate at a stable steady state. \\
The second option to ensure stability of the system arises from the negative feedback of ATP
upon the combined PFK-HK reaction. 
Fig. \ref{fig:EV_histogram}b shows the distribution of the 
largest real part $\lambda^\mathrm{max}_R$ 
of the eigenvalues for a nonzero feedback strength~$\xi>0$.
Again, the distribution is markedly shifted towards negative values, increasing the
probability of a stable steady state. \\
In more detail, Fig.~\ref{fig:EV_Stability} depicts the distribution of $\lambda^\mathrm{max}_R$ 
as a function of the feedback strength $\xi$.
As can be observed, in the absence of the regulatory feedback the system is prone to instability,
i.e. it is not possible (or rather unlikely) for the observed operating point to exist as a stable
steady state. 
Subsequently, as the feedback strength is increased, the probability of obtaining a stable steady state increases.
For an intermediate value $\xi=1$ the system is fully stable: Any realization of
the Jacobian will result in a stable steady state, independent of the detailed functional form of
the rate equations or their associated parameters.    
However, as the feedback is increased further, the operating point again looses its stability.
This time the instability arises out of a Hopf bifurcation, indicating the presence
of sustained oscillations. \\
Based on these findings, we can summarize some essential properties 
of the pathway depicted in Fig.~\ref{fig:Glycolysis}: 
Given the experimentally observed metabolite concentrations and flux values, 
our results show that
in the absence of the assumed regulatory interaction 
it would not be possible (or, at least highly unlikely) to observe either 
sustained oscillations or a stable steady state.
However, for sufficiently large inhibitory feedback, the system will inevitably
exhibit sustained oscillations. 
Furthermore, as the feedback strength $\xi \in [0,n]$ is bounded by the  
Hill coefficient $n$ of the (unspecified) rate equation, $n \geq 2$ is required for the existence of sustained oscillations. \\
We emphasize again that these results do not rely on any explicit kinetic model 
of the system. 
As demonstrated, our method allows to derive the likeliness or plausibility of 
the experimentally observed oscillations, as well as the specific 
kinetic requirements for oscillations to occur, 
without referring to the detailed functional form of the rate equations.

\section*{The photosynthetic Calvin Cycle}
\begin{figure}
\centering{\includegraphics[width=0.45\textwidth]{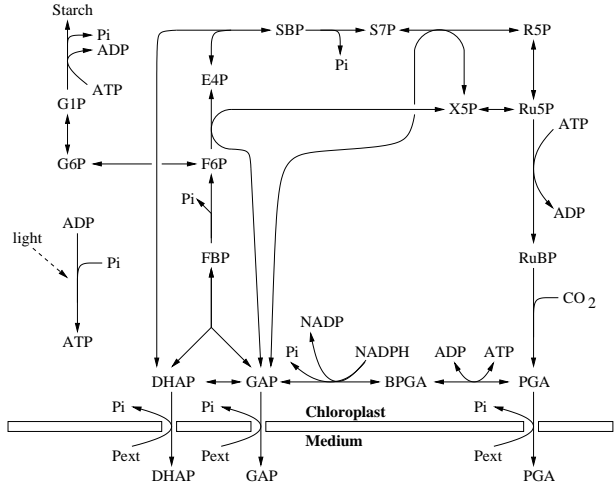}}
 \caption{\label{fig:CalvinCycle} \small
The photosynthetic Calvin Cycle, adapted from earlier kinetic models~\cite{PRP88,POL01}.
The systems consists of $18$ metabolites, subject to two conservation relations,
and $20$ reactions, including three export reactions, starch synthesis and regeneration of ATP.
The rank of the stoichiometric matrix is $\mathrm{rank}(\mathbf{N})=16$, leaving $4$ independent
steady state reaction rates.
Throughout this section, the steady state concentrations and flux values are as 
reported by Petterson and Ryde-Petterson~\cite{PRP88},
describing the pathway under conditions of light and CO$_2$ saturation.
For metabolite abbreviations see the Supplementary Information.}
\end{figure}

\begin{figure}
\centering{\includegraphics[width=0.3\textwidth]{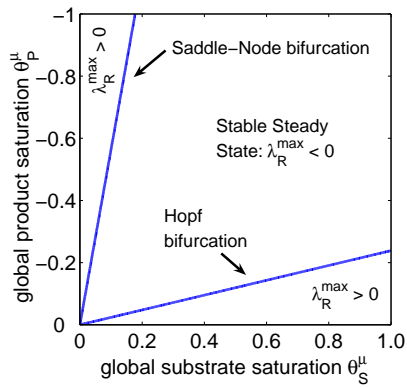}}
 \caption{\label{fig:CalvinBifurcation} \small
The bifurcation diagram of the Calvin cycle at
the observed operating point with respect to the two global saturation 
parameters~$\theta^{\mu}_S \in (0,1]$ and $\theta^{\mu}_P \in [0,-1]$.   
}
\end{figure}

\begin{figure}
\centering{\includegraphics[width=0.5\textwidth]{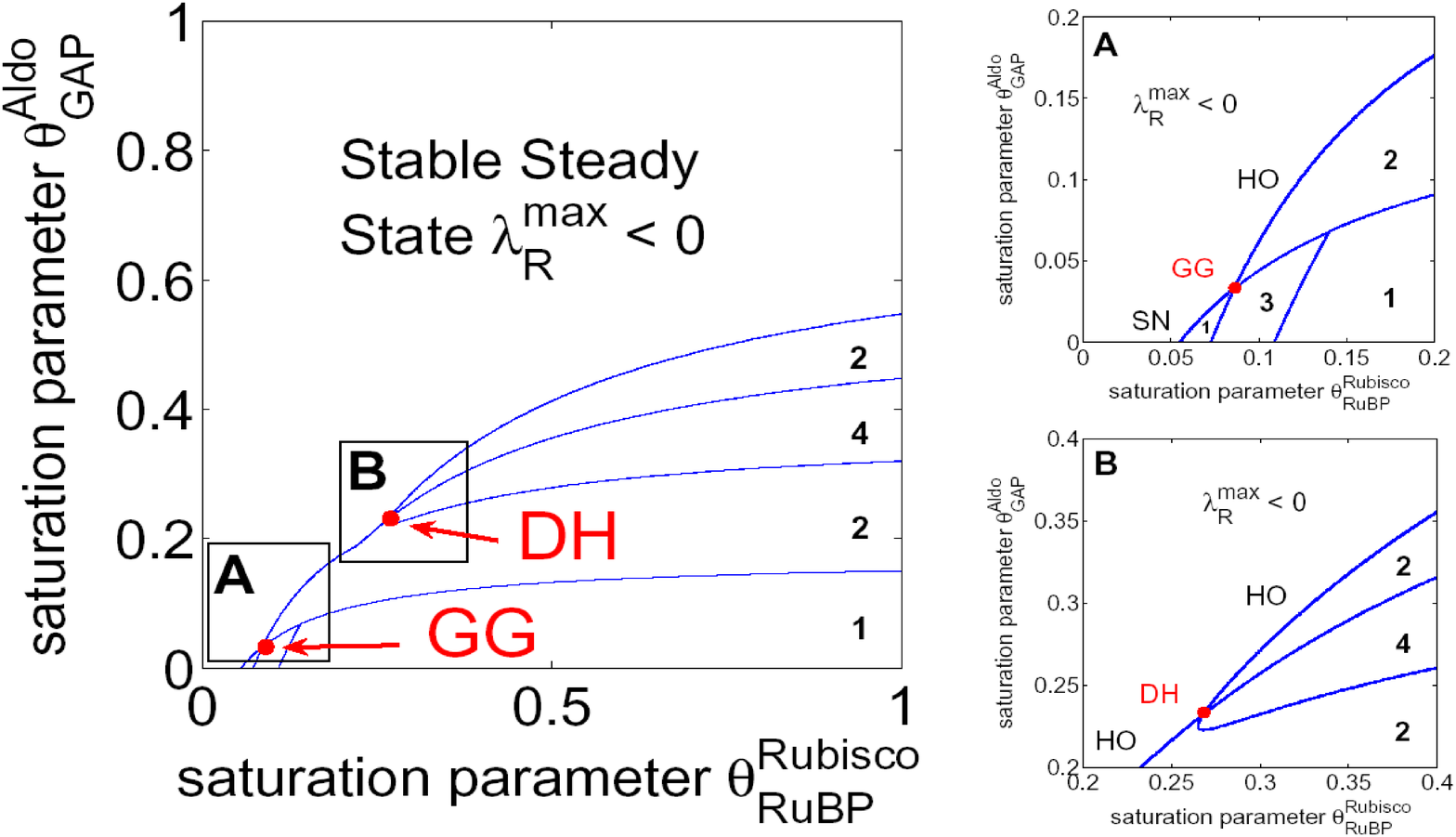}}
\caption{\label{fig:Calvin_Bifdiagram} \small
The bifurcation diagram of the Calvin cycle as a function of the saturation of the rubisco
reaction with respect to RuBP and the saturation of the Aldolase reaction with respect to GAP,
while all other saturation parameters are fixed to specific values. Bifurcation lines are depicted in blue, 
numerals indicate the number the positive real parts within the spectrum of eigenvalues. 
{\em Inset A:} The system gives rise to a  Gavrilov-Guckenheimer (GG) bifurcation, formed by the interaction
of a Hopf (HO) and a saddle-node (SN) bifurcation.
{\em Inset B:} The interaction of two Hopf (HO) bifurcations gives rise to a double Hopf (DH) bifurcation.
}
\end{figure}

The CO$_2$ assimilating Calvin Cycle, taking place in the chloroplast stroma of plants, 
is a primary source of carbon for all organisms and of central  
importance for many biotechnological applications. 
However, even when restricting an analysis to the core pathway shown in Fig.~\ref{fig:CalvinCycle},
the construction of a detailed kinetic model already entails considerable challenges with respect 
to the required rate equations and kinetic parameters~\cite{PRP88,POL01}. \\
In the following, we thus use the scheme depicted in Fig.~\ref{fig:CalvinCycle} to
demonstrate the applicability of our approach to a system of a reasonable complexity.  
In particular, we seek to describe a general strategy to extract information about the dynamical capabilities of the system, without
referring to an explicit set of differential equations.
Our agenda focuses on 
{\em (i)} the stability and robustness of the experimentally observed operating point,
{\em (ii)} the relative impact or importance of each reaction upon the dynamical properties of the system,
{\em (iii)} the existence and quantification of different dynamical regimes, 
such as oscillations and multistability, and 
{\em (iv)} the possibility of complex or chaotic temporal behavior. \\
Starting point is again a particular observed state, characterized by the 
vector of metabolite concentrations $\mathbf{S}^\mathbf{0}$ and flux values~$\bm{\nu}^\mathbf{0}$. 
Although additional knowledge on the reactions is often available,
for the moment we assume that all reactions depend
only on their substrates and products, with parameters~$\theta^{\mu}_S \in (0,1]$ and 
$\theta^{\mu}_P \in [0,-1]$, respectively.   
This information, embedded within the matrices~$\bm{\Lambda}$ and $\bm{\theta}_\mathbf{S}^\mathbf{\bm \mu}$,
constitutes the structural kinetic model of the Calvin cycle at the observed operating point. \\ 
As a first approximation, we commence with global saturation parameters, $\theta^{\mu}_S$ and $\theta^{\mu}_P$,
set equal for all reactions.
Though clearly oversimplified, the resulting bifurcation diagram, 
depicted in Fig.~\ref{fig:CalvinBifurcation},
already reveals some fundamental dynamical properties of the system:
First, the observed operating point is indeed a 
stable steady state for most parameters $\theta^{\mu}_S$ and $\theta^{\mu}_P$. 
Interestingly, however, in the absence of product inhibition $\theta^{\mu}_P=0$,  a steady state is no longer feasible. 
In particular, for pure irreversible mass-action kinetics ($\theta^{\mu}_S=1$, $\theta^{\mu}_P=0$), 
corresponding to a non-enzymatic chemical system, the pathway could not operate at the observed
steady state. 
Second, for low product saturation ($\theta^{\mu}_P$ close to zero), a Hopf bifurcation occurs. 
While this does not necessarily imply that this region within parameter space is 
actually accessible under normal physiological conditions,
it shows the dynamical capability of the system to generate sustained oscillations, i.e. there
exists a region in parameter space that allows for oscillatory behavior. 
Additionally, for low values of the substrate saturation $\theta^{\mu}_S$, a saddle-node bifurcation occurs.
This shows that the observed steady state will eventually loose its stability, i.e. there are 
conditions under which the observed steady state is no longer stable. 
Noteworthy, both dynamical features have been 
observed for the Calvin cycle: 
Photosynthetic oscillations are known for many decades
and have been subject to extensive experimental and numerical studies~\cite{RP91}. 
Furthermore, multistability was recently found in a detailed kinetic model of the 
Calvin cycle and verified {\em in vivo}~\cite{POL01}. \\
To proceed with a systematic analysis, the next step is to drop the assumption of
global saturation parameters. 
All individual parameters $\theta^{\mu}_S \in (0,1]$ are now allowed
to take arbitrary values in the unit interval, reflecting the full spectrum of
possible dynamical capabilities of the metabolic system.    
For simplicity, all reactions are still restricted to weak saturation by their products~$\theta^{\mu}_P = 1/3$. 
Of foremost interest is again the robustness of the experimentally observed operating point:
Evaluating an ensemble of $5 \cdot 10^5$ random realizations of the Jacobian at this operating 
point, the system gives rise to a stable steady state in $\approx 94.3\%$ of all cases (see Supplementary
Information for convergence and dependence on ensemble size).
Thus the stability of the observed operating point is indeed generic and does not
rely on a specific choice of the kinetic parameters. \\
As for the remaining $\approx 5.7\%$ of models, corresponding to the case 
where the observed operating point is instable, about $5.1\%$ give
rise to a single positive eigenvalue. Only $\approx 0.6\%$ correspond to a more
complex situation, with two or more real parts larger than zero. 
The latter case, though only restricted to a small region within parameter space,
holds profound implications for the possible dynamics of the system. 
As a further step within our approach, 
the existence of certain bifurcations of higher codimension allows to
predict the possibility of specific dynamics (see {\em Materials and Methods}).  
Fig.~\ref{fig:Calvin_Bifdiagram} shows a bifurcation diagram of
the Calvin cycle within a particular region of parameter space where such bifurcations occur.  
Here, the system gives rise to a Gavrilov-Guckenheimer (GG) bifurcation, implying
the existence of quasiperiodic dynamics and making the existence of chaotic dynamics likely. 
In close vicinity of the GG bifurcation, we also find a double Hopf (DH) bifurcation, 
formed by the interaction of two codimension-1 Hopf bifurcations.  
The generic existence of a chaotic parameter region close to the DH bifurcation can
be proven~\cite{Kuz95,GEF05}.  \\
Thus our results demonstrate the possibility of quasiperiodic 
and chaotic dynamics for the model of the photosynthetic Calvin cycle shown in Fig.~\ref{fig:CalvinCycle}, 
without relying on any particular assumptions about the functional form of the kinetic rate equations. 
Furthermore, being a quantitative method, we can assert that complex dynamics at the 
operating point are confined to a rather small region in parameter space and that the experimentally
observed steady state is generically stable.

\section*{Discussion and Conclusions}
We have presented a systematic approach to explore and quantify 
the dynamic capabilities of a metabolic system, 
without requiring to specify the detailed functional form of 
any of the involved rate equations.
Starting with a parametric representation of the Jacobian matrix, 
constructed in such a way that each element is either directly experimentally
accessible or amenable to a clear biochemical interpretation,
we look for characteristic bifurcations that give insight into the possible dynamics of the system.
Our method then builds upon the construction of a large ensemble of models, 
encompassing {\em all possible} explicit kinetic models,  
to statistically explore and quantify the parameter region associated 
with a specific dynamical behavior. \\
One of the primary advantages is that all results relate to
a particular experimentally observed operating point of the system.
In this respect, the method contrasts with the trivial alternative 
of drawing all (known) nonzero elements of the Jacobian from a random distribution.   
While the latter would likewise allow to indicate e.g. the possibility of oscillatory behavior,
it fails to actually quantify the associated parameter region at a particular observed state. 
Only by means of our parametric representation of the system, we are in the position to
identify crucial reaction steps that predominantly contribute to
the stability, and thus robustness, of an experimentally observed state and can give 
explicit biochemical conditions
for which a specific dynamical behavior can be expected. 
Furthermore, by taking bifurcations of higher codimension into account, we go
beyond the usually considered case and are able to predict the 
possibility of complex or chaotic dynamics - often 
a nontrivial task, even if an explicit kinetic model is available. \\
We emphasize that our approach is not restricted to an analysis of the
bifurcations and stability properties of metabolic systems.
Once the parametric representation of the Jacobian is obtained, 
it can serve a multitude of purposes.
The Jacobian holds a wealth of information, including the systems response to (small) perturbation, 
the hierarchy of characteristic timescales~(Modal Analysis)~\cite{HS96},
as well as the possibility to deduce the flux and concentration control coefficients, 
defined in the realm of Metabolic Control Analysis~\cite{HS96}.   
Along similar lines, it is thus possible to explore the influence 
of particular reactions and their associated saturation parameters
upon more general features of the system.
In this respect, structural kinetic modeling also serves as a prequel to
explicit mathematical modeling, aiming to identify 
crucial reaction steps and parameters in best time.

\appendix
\section*{\sc Materials and Methods}
{\small
\subsection*{\sc The Interpretation of the Saturation Matrix}
Our approach relies crucially on the interpretation of the elements of 
the matrix $\bm{\theta}_\mathbf{x}^\mathbf{\bm \mu}$.
As a simple example, consider a single bilinear 
reaction rate of the form $\nu(S_1,S_2)=v_\mathrm{max}S_1 S_2$.
Then, according to Eq.~(\ref{eq:rates}), the normalized rate is $\mu(x_1,x_2)=x_1 x_2$, thus
\begin{equation}
\theta_{x_i}^{\mu} = 
\left. \frac{\partial\mu}{\partial x_1} \right|_{\mathbf{x}^0=\mathbf{1}} =
\left. \frac{\partial\mu}{\partial x_2} \right|_{\mathbf{x}^0=\mathbf{1}} = 1 
~~.
\end{equation} 
In the case of Michaelis-Menten kinetics $\nu(S)=v_\mathrm{max}S/(K_\mathrm{M}+S)$, 
depending on a single substrate $S$, we obtain 
\begin{equation}
\mu(x) = x \, \frac{K_M+S^0}{K_M + x S^0} \quad \Rightarrow \quad
\theta_{x}^{\mu} 
= \frac{1}{1+S^0/K_\mathrm{M}} \in [0,1]
\end{equation}
Clearly, the partial derivative $\theta_{x}^{\mu} \in [0,1]$ measures the degree of saturation or, likewise, 
the effective order of the reaction at the steady state~$S^0$.
The limiting cases are $\lim_{S^0\rightarrow 0}\theta_{x}^{\mu}=1$ (linear regime)
and $\lim_{S^0\rightarrow\infty}\theta_{x}^{\mu}=0$ (full saturation). 
This implies that the saturation parameter indeed covers the full interval, 
which holds likewise for the general case of Eq.~(\ref{eq:GeneralRate}).
For additional instances of specific rate functions, as well as a proof of Eq.~(\ref{eq:GeneralRate}), 
see the Supplementary Information. \\
Note that, except for the change in variables, the 
saturation parameters $\bm{\theta}_\mathbf{x}^\mathbf{\bm \mu}$  
are reminiscent of the scaled elasticity coefficients, as defined in the 
realm of Metabolic Control Analysis~\cite{HS96}.
However, for our reasoning to hold, the analysis is restricted to unidirectional reactions, i.e.
in the case of reversible reactions, forward and backward terms have to be treated separately. 
As the denominator is usually preserved for both terms, this does 
not give rise to additional free saturation parameters. \\
Another close analogy to the saturation parameters is found 
within the power-law approximation~\cite{HS96},
where each enzyme kinetic rate law is replaced by a 
function of the form $\nu_j(\mathbf{S})=\alpha_j \prod_i S_i^{g_{ij}}$.
In fact, the power-law formalism can be regarded as the simplest 
possible way to specify explicit nonlinear functions that are consistent with a given Jacobian.
Applying the transformation of Eq.~(\ref{eq:rates}), we obtain $\mu_j(\mathbf{x}) = \prod_i x_i^{g_{ij}}$,
thus $\theta_{x_i}^{\mu_j} = g_{ij}$.
However, beyond the properties of the Jacobian itself, only little confidence can
be placed in an actual numerical integration of these functions~\cite{HS96}. 
Generally, it is possible to specify several classes of explicit functions that, by construction,
result in a given Jacobian, but have no, or only little, biochemical justification otherwise.   
Consequently, within our approach, we opt for utilizing the properties of the parametric representation
of the Jacobian directly, instead of going the loop way via auxiliary ad-hoc functions.

\subsection*{\sc Dynamics and Bifurcations}
One of the foundations of our approach is the fact that 
knowledge of the Jacobian matrix alone is sufficient to deduce certain
characteristic bifurcations of a metabolic system. 
In general, the stability of a steady state is lost either in a {\em Hopf bifurcation} (HO) 
or in a bifurcation of {\em saddle-node} (SN) type, both of codimension-1. 
Of particular interest to reveal insights about the dynamical behavior of systems are
also bifurcations of higher codimension, such as the 
{\em Takens-Bogdanov} (TB), the {\em Gavrilov-Guckenheimer} (GG) 
and the {\em double Hopf} (DH) bifurcation~\cite{Gross04,Kuz95}. 
Each of these local bifurcations of codimension-2 arises out of
an interaction of two codimension-1 bifurcations and 
has important implications for the possible dynamical behavior.  
For instance the TB bifurcation 
indicates the presence of a homoclinic bifurcation 
and therefore the possibility of spiking or bursting behavior. 
The presence of a Gavrilov-Guckenheimer bifurcation shows that complex (quasiperiodic or chaotic) 
dynamics exist generically in a certain parameter space. 
In the same way the double Hopf bifurcation indicates 
the generic existence of a chaotic parameter region. 
For details see~\cite{Gross04,Kuz95}
and the {\em Supplementary Information}.

}

\end{document}